\documentclass[preprints,article,accept,moreauthors,pdftex]{Definitions/mdpi}

\firstpage{1} 
\pubvolume{12}
\issuenum{3}
\articlenumber{404}
\pubyear{2022}
\copyrightyear{2022}
\datereceived{} 
\dateaccepted{} 
\datepublished{} 
\hreflink{https://doi.org/10.3390/life12030404} 
\pdfoutput=1

\usepackage[version=4]{mhchem}
\usepackage{siunitx}
\DeclareSIUnit\year{yr}
\DeclareSIUnit\bar{bar}

\Title{Possible Ribose Synthesis in Carbonaceous Planetesimals} 

\TitleCitation{Possible Ribose Synthesis in Carbonaceous Planetesimals}

\Author{Klaus Paschek $^{1,}$*\orcidA{}, Kai Kohler $^{1,2}$, Ben K. D. Pearce $^{3,4,\dagger}$\orcidB{}, Kevin Lange $^{5,\ddagger}$\orcidC{}, Thomas K. Henning $^{1}$\orcidD{}, Oliver~Trapp~$^{1,2}$\orcidE{}, Ralph E. Pudritz $^{3,4}$\orcidG{} and Dmitry A. Semenov $^{1,2}$\orcidH{}}

\AuthorNames{Klaus Paschek, Kai Kohler, Ben K. D. Pearce, Kevin Lange, Thomas K. Henning, Oliver Trapp, Ralph E. Pudritz and Dmitry A. Semenov}

\AuthorCitation{Paschek, K.; Kohler, K.; Pearce, B.K.D.; Lange, K.; Henning, T.K.; Trapp, O.; Pudritz, R.E.; Semenov, D.A.}

\address{%
$^{1}$ \quad Max Planck Institute for Astronomy, K{\"o}nigstuhl 17, 69117 Heidelberg, Germany;\linebreak kai.kohler@cup.uni-muenchen.de (K.K.); henning@mpia.de (T.K.H.);\linebreak oliver.trapp@cup.uni-muenchen.de (O.T.); semenov@mpia.de (D.A.S.)\\
$^{2}$ \quad Department of Chemistry, Ludwig-Maximilians-University Munich, Butenandtstr.~5-13, House F, \mbox{81377 Munich, Germany} \\
$^{3}$ \quad Origins Institute, McMaster University, ABB, 1280 Main Street West, Hamilton, ON, L8S 4M1, Canada; bpearce6@jhu.edu (B.K.D.P.); pudritz@physics.mcmaster.ca (R.E.P.)\\
$^{4}$ \quad Department of Physics and Astronomy, McMaster University, ABB, 1280 Main Street West, \mbox{Hamilton, ON, L8S 4M1, Canada}\\
$^{5}$ \quad Institute for Theoretical Astrophysics, Center for Astronomy, Heidelberg University, Albert-Ueberle-Str.~2, 69120 Heidelberg, Germany; k.lange@uva.nl}

\corres{Correspondence: paschek@mpia.de}

\firstnote{Current address: Department of Earth and Planetary Science, Johns Hopkins University, \mbox{3300 San Martin Drive,} Baltimore, MD 21218, USA.} 
\secondnote{Current address: Anton Pannekoek Institute for Astronomy, University of Amsterdam, Science-Park 904, \mbox{1098 XH Amsterdam, The Netherlands.}}

\abstract{The origin of life might be sparked by the polymerization of the first RNA molecules in Darwinian ponds during wet-dry cycles. The key life-building block ribose was found in carbonaceous chondrites. Its exogenous delivery onto the Hadean Earth could be a crucial step toward the emergence of the RNA world. Here, we investigate the formation of ribose through a simplified version of the formose reaction inside carbonaceous chondrite parent bodies. Following up on our previous studies regarding nucleobases with the same coupled physico-chemical model, we calculate the abundance of ribose within planetesimals of different sizes and heating histories. We perform laboratory experiments using catalysts present in carbonaceous chondrites to infer the yield of ribose among all pentoses (5Cs) forming during the formose reaction. These laboratory yields are used to tune our theoretical model that can only predict the total abundance of 5Cs. We found that the calculated abundances of ribose were similar to the ones measured in carbonaceous chondrites. We discuss the possibilities of chemical decomposition and preservation of ribose and derived constraints on time and location in planetesimals. In conclusion, the aqueous formose reaction might produce most of the ribose in carbonaceous chondrites. Together with our previous studies on nucleobases, we found that life-building blocks of the RNA world could be synthesized inside parent bodies and later delivered onto the early Earth.}


\begin{document}

\section{Introduction}\label{sec:intro}

Ribose was recently identified in the soluble organic matter of carbonaceous chondrites~\cite{Furukawa2019}, with measured concentrations of \SIrange{4.5}{25}{ppb} (parts per billion). Along with comets, the parent bodies of carbonaceous chondrites or their fragments are considered as one of the major sources of pristine organic matter that have been exogenously delivered onto the Hadean and Eoarchean Earth during the late heavy bombardment, see, e.g., \cite{Chyba1990,Chyba:1992bp,Gomes2005,Kooten2021,Fischer-Godde2020,Pizzarello2017}, and are still falling to Earth nowadays.

One of the favored theories of the origin of life relies on the finding that RNA molecules can act as both catalysts of self-replication and genetic information storage, see, e.g.,~\mbox{\cite{Rich1962,Gilbert1986,Kruger1982,Guerrier-Takada1983,Guerrier-Takada1984,Zaug1986,Cech1986,Johnston2001,Vaidya2012}.} So-called ribozymes allow solving the fundamental ``chicken-or-the-egg'' problem in the process of abiogenesis and the emergence of life: What came first, the genetic information macromolecules like RNA and DNA coding for protein sequences or the enzymatic proteins catalyzing the formation of information storage molecules and other biomolecular reactions? One mutually compatible molecule is needed for the synthesis and \textit{survival} of the other one, and vice versa.

RNA polymers could solve this dilemma by providing both these functionalities, serving as the universal ``Swiss Army knife'' to the organocatalysis of the first complex biomolecules. The polymerization of oligonucleotides and RNA-like molecules was shown to be possible in the hydrothermal field settings, in the presence of clays or salts \cite{Ferris1996,Ferris2004,DaSilva2015}, metal ion catalysts \cite{Orgel2004}, and in lipid bilayers \cite{Toppozini2013}. In particular, wet-dry cycles, which were meant to simulate the natural cycling in warm little ponds (WLPs) on freshly formed volcanic islands on the Hadean Earth, were shown to promote the formation of RNA-like chains made up of up to 300 linked residues \cite{DaSilva2015,Damer2020}. \citet{Becker2018,Becker2019} showed in lab experiments that the formation of RNA nucleosides and nucleotides was plausible in \textit{one pot} in WLPs during wet-dry cycles (requiring specific and changing conditions with respect to temperature and pH). They added ribose without forming it in the same pot, which makes the possibility of an exogenous delivery by meteorites an intriguing hypothesis. The different pathways to nucleosides and nucleotides shown in experiments by Orgel and colleagues \cite{Fuller1972a,Fuller1972b}, \citet{Kim2017,Kim2018}, and \citet{Nam2018} also all required the prior presence of ribose. \citet{Pearce2017} considered the delivery of nucleobases by carbonaceous chondrites to WLPs as a source for the formation and polymerization of oligonucleotides during wet-dry cycles. Similarly, ribose-rich meteorites might provide another crucial ingredient for this process that lead to the emergence of the first RNA-like molecules on the early Earth.

The crucial components that make up the backbone of these linked oligonucleotides are the pentose (5C), ribose, and phosphate groups. Our study sought to model the formose reaction pathway leading to the synthesis of ribose inside carbonaceous planetesimals. We used the same model as developed in our previous study by \citet{Paschek2021}, where we studied the formation of nucleobases inside the parent bodies of carbonaceous chondrites. Our model comprised an up-to-date thermochemical equilibrium model coupled with a 1D thermodynamic planetesimal model to calculate the abundances of prebiotic molecules via abiotic synthesis pathways under realistic conditions inside parent body planetesimals. By applying the same model to ribose, we elucidated a more comprehensive understanding of the formation of crucial building blocks of the RNA world in outer space, very early in the history of our solar system. 

The theoretical background is provided in Section~\ref{sec:theory}. First, we introduce our approach to model the formose reaction pathway, focusing on the synthesis of ribose. Further, we briefly recall our new concept to estimate the initial concentrations of the volatiles in carbonaceous planetesimals, which was explained in more detail in \citet{Paschek2021}. Next, in Section~\ref{sec:computations}, we outline our computational methods. We also explain how the Gibbs free energy of formation of glycolaldehyde needed for our theoretical model was estimated, as it is missing in the database that we used. In Section~\ref{sec:results}, we first present our experimental results regarding the efficiency of ribose formation among other 5Cs in the presence of various mineral catalysts representative of carbonaceous chondrites. Second, we present our theoretical results, incorporating the ribose yields obtained in our experiments. We analyze our results and compare them to the measured values in carbonaceous chondrites. The discussion and conclusions follow in Section~\ref{sec:discussion}. We discuss the possible decomposition processes of ribose and gave a suggestion for the region within the planetesimals with the likely highest ribose abundance.

\section{Materials and Methods}

\subsection{Theory}\label{sec:theory}

\subsubsection{Formose Reaction Pathway}\label{sec:reactions}

The formose reaction, first described by \citet{Butlerow1861}, is a reaction network forming a variety of sugars. It starts with formaldehyde in an aqueous solution. In a self-condensation, a dimerization of formaldehyde leads to the formation of glycolaldehyde, starting an autocatalytic cycle \cite{Breslow1959}. The formation of glycolaldehyde itself is not yet fully understood, since a freshly distilled formaldehyde solution leads only to the nonproductive reaction to methanol and formate \cite{Cannizzaro1853}. Small amounts of sugars or impurities are needed to start the formose reaction, e.g., \SI{3}{ppm} (parts per million) of glycolaldehyde are sufficient \cite{Socha1980}. Recent experimental and numerical studies suggested that glycolaldehyde could be formed in interstellar clouds, either by surface hydrogenation of \ce{CO} molecules on icy dust grains~\cite{Fedoseev2015}, or by formaldehyde reacting with its isomer hydroxymethylene \cite{Eckhardt2018}. Glycolaldehyde drives the formose reaction toward more complex sugars. For this reason, we started our laboratory experiments with a solution of formaldehyde and glycolaldehyde and modeled the reaction accordingly in our theoretical studies.

A multitude of different catalysts is effective in the formose reaction. Very effective catalysts are hydroxides, carbonates, and oxides of alkali/alkaline earth metals, as well as aluminosilicates, tertiary amines, lanthanide hydroxides, thallium hydroxide, and lead oxide, see, e.g., \cite{Iqbal2012}. In particular, hydroxides and carbonates are of great interest, as these are commonly found in carbonaceous chondrites, see, e.g., \cite{Barber1981}.
In the context of prebiotic chemistry, a presumed reaction pathway for the synthesis of molecules involved in the emergence of life has to start from abiotic and naturally abundant molecules. 
Our considered reaction pathway for the formation of ribose (\ce{C5H10O5}) from formaldehyde (\ce{CH2O}) and glycolaldehyde (\ce{C2H4O2}) can be summarized as
\begin{equation}\label{equ:formose}
    \ce{CH2O{}_{(aq)} + 2C2H4O2{}_{(aq)} ->[catalyst] C5H10O5{}_{(aq)}}.
\end{equation}

Here, ``catalyst'' stands for either one or the other of the following hydroxides or carbonates: \ce{Ca(OH)2}, \ce{CaCO3}, \ce{KOH}, or \ce{K2CO3}. We used one of these catalysts in each run of our laboratory experiments. The formose reaction in Equation~\eqref{equ:formose} is an oversimplification of the far more complex reaction network toward sugars. Hence, we performed the laboratory work to correct our theoretical results by using realistic ribose yields measured in our experiments.

In the framework of the previous studies of \citet{Pearce2015,Pearce2016} and \citet{Paschek2021}, an ID-number was assigned (maximum two digits) to each considered reaction pathway (there reaction pathways for the formation of nucleobases). We extended this numbering scheme to the formose reaction and assigned the no.~101 to the reaction pathway considered here in Equation~\eqref{equ:formose} (we reserved no.~100 for the potential formation of ribose from formaldehyde only, without initially present glycolaldehyde, which was not considered here).

Theoretical and experimental studies proposed other possible reactants and reaction pathways for the formation of ribose and RNA precursors in combination with nucleobases. \citet{Weber2006} demonstrated a stereospecific formation of tetroses from glycolaldehyde catalyzed by peptides. This could be one of the possible solutions to the unsolved question of how the homochirality of life emerged. \citet{Jeilani2020} used quantum chemistry calculations to verify the possibility of abiotic ribose and RNA nucleoside synthesis by free radical reactions. They started from formaldehyde and used \ce{Ca^{2+}} and \ce{CaOH+} cations as catalysts. The hydroxymethyl radical \ce{^{.}CH2OH} was identified as a potential intermediate in the dimerization of formaldehyde and the autocatalytic cycle. \mbox{\citet{Teichert2019}} and \citet{Kruse2020} presented a new direct formation pathway to DNA nucleosides. They started with nucleobases and specifically formed deoxyribose by condensation with acetaldehyde and sugar-forming precursors. \citet{Saladino2015} and \citet{Sponer2016} showed the formation of nucleosides starting from formamide catalyzed by powdered meteorites. Additionally, \citet{Eschenmoser2007a,Eschenmoser2007b} proposed a hypothetical reaction pathway starting from \ce{HCN}. Amino acids and carbohydrates might be formed with glyoxylate (glyoxylic acid in neutral solution) and its dimer dihydroxyfumarate as intermediates in the so-called ``glyoxylate scenario''. \citet{Banfalvi2021} reviewed this scenario in comparison to the formose reaction. Further reviewed aspects are an alternative mechanism for the formation of RNA starting with the ribose-phosphate backbone, which then binds nucleobases, skipping ribonucleotides as intermediate molecules. Moreover, this study described why ribose is the best fitting aldopentose for the build-up of RNA, as ribose allows for the maximum flexibility of RNA.

\subsubsection{Initial Concentrations of Reactants}\label{sec:concs}

In order to model the formose reaction in Equation~\eqref{equ:formose} with our theoretical chemical model, we used the initial concentrations of the reactants as inputs to determine the resulting abundance of ribose. Comets are believed to have the most pristine composition, and to most closely reflect the conditions that prevailed before or during the early stages in the formation of our solar system. Therefore, comets are the only reservoir of such pristine objects in the solar system that still exists and is accessible to measurements. We took the abundances measured spectroscopically in comets \cite{Mumma2011} (and references therein) as the first reference values. With this, we followed the same approach as described in the previous studies by \citet{Cobb2015} and \citet{Pearce2016}.

Nevertheless, the icy pebbles making up the source material of the parent bodies of carbonaceous chondrites are believed to originate from more inner and warmer regions further inside the solar nebula than those of comets. The main-belt asteroid 19~Fortuna, orbiting the Sun at $\sim$$\SI{2.5}{au}$, was identified as the potential parent body source of CM (Mighei-type) meteorites \cite{Burbine2002}. Therefore, we assumed the inner region of \SIrange{2}{3}{au} to be the forming location of carbonaceous chondrite parent bodies in the solar nebula (in a first approximation neglecting possible radial migration processes). This region was more distant to the proto-Sun than the water snowline at ${T\lesssim\SI{150}{\kelvin}}$ \cite{1977E&PSL..36....1W,2001Sci...293...64A,2001M&PS...36..671C,Lodders03,2005PNAS..10213755B,2018GeCoA.239...17B,2020ApJ...897...82V,Oberg_Bergin21,Lichtenberg2021}. As a result, water ice was supposed to be preserved in the source material of carbonaceous chondrites. On the other hand, this region is inside the sublimation zone of the reactant formaldehyde \cite{1977E&PSL..36....1W,2001Sci...293...64A,2001M&PS...36..671C,Lodders03,2005PNAS..10213755B,2018GeCoA.239...17B,2020ApJ...897...82V,Oberg_Bergin21,Lichtenberg2021}, which has a sublimation temperature of ${\sim}$$\text{\SIrange{40}{45}{\kelvin}}$ \cite{Cuppen_ea17}. Consequently, the icy pebbles in this region are expected to lose a substantial fraction of the most volatile constituents, e.g., formaldehyde, which diffused through the monolayers of water ice and sublimed into space, leading to a more volatile-poor water ice mantle \cite{C5CP00558B}. Thus, the more volatile formaldehyde should have been depleted in the carbonaceous chondrites' building blocks compared to the pristine ices in the comet-forming zone. 

It was predicted that carbonaceous planetesimals were rapidly assembled via streaming instabilities from the source material \cite{Johansen_ea07,Ormel2010,Klahr_Schreiber20}. However, unlike comets, these pristine pebbles were not preserved. Therefore, we had to refer to models predicting the depletion of formaldehyde \cite{C5CP00558B} to simulate the physico-chemical processes in the whole solar nebula~\cite{Visser_ea09,SW11,Drozdovskaya_ea16,Bergner_Ciesla21,Lichtenberg2021}. We then compared the remaining volatile abundances between the forming regions of comets and carbonaceous chondrite parent bodies in these solar nebular models. \citet{Drozdovskaya_ea16} simulated two different scenarios of the solar nebula collapsing into a protoplanetary disc. They predicted a depletion of formaldehyde ice by about three or more orders of magnitude when comparing between regions at \SIrange{1}{10}{au} and ${>}$$\SI{30}{au}$. The abundance changed between outer and inner regions by a factor of ${\le}$$\num{4.25e-3}$ (compare to Table~4 in their publication). In the models of \citet{Visser_ea09}, \citet{SW11}, and \citet{Bergner_Ciesla21} similar or even higher depletion factors were predicted. The large range of possible depletion factors results from different model assumptions, different computed scenarios, considered physico-chemical mechanisms, and large uncertainties within and between the models. Note that the water ice was not substantially depleted at \SIrange{2}{3}{au} in all the models, which coincides with the considerations about the water snowline mentioned above. This allowed us to normalize all molecular abundances to that of water~ice. 

In the temperature programmed desorption (TPD) experiments in the laboratory, it was shown that the sublimed volatiles were not able to leave the water ice matrix freely. A fraction of the volatile ices remained trapped in the water ice and co-desorbed at higher temperatures of ${T \gtrsim \SI{150}{\kelvin}}$ when the water ice sublimed. However, icy pebbles in the solar nebula might have gradually lost their volatile content over thousands of years and became more volatile-poor compared to the results of the TPD experiments, which were conducted over short timescales (hours--days) \citep{Cuppen_ea17,Potapov_McCoustra21}. Another TPD experiment showed that some of the formaldehyde polymerizes in reaction with water to polyoxymethylene and therefore did not co-desorb with water at high temperatures \cite{Noble2012}. Polymerization of formaldehyde could reduce the amount of its monomer form that was available to the synthesis of ribose in the formose reaction. When the formed parent body planetesimal heated up, a fraction of the formaldehyde could be trapped in its polymeric form, preventing it from participating in the synthesis of more complex organics.

In summary, we used a factor of $10^{-3}$ to reduce the abundance of formaldehyde measured in comets, corresponding to a conservative estimate of an upper bound motivated by the value of ${\le}$$\num{4.25e-3}$, given in the solar nebula model by \citet{Drozdovskaya_ea16}. This value was chosen in this approximate manner since it was based on many assumptions and uncertainties. As the solar nebula models are in general strong simplifications of the different complex desorption, trapping, and chemical mechanisms, which were observed in, and deduced from, the TPD experiments and could have occurred during the formation of the protoplanetary disc of the solar system, they most likely overestimate the depletion of volatiles.

Moreover, different assumed sizes of the icy pebbles forming the carbonaceous planetesimals introduce largely different predictions for the diffusion times of formaldehyde and other volatiles, since the molecules can leave faster through the porous water ice layers in smaller pebbles. The diffusion rate of formaldehyde at ${T \gtrsim \SI{90}{\kelvin}}$ was measured experimentally and predicted via molecular dynamics calculations \cite{C5CP00558B} (and references therein). The resulting diffusion and depletion timescales of volatiles then depend on the thickness of the bulk water ice that needs to be passed, and on the sizes of the icy pebbles accordingly. Further, uncertainties are caused by possible migration processes within the solar nebula and the protoplanetary disc \cite{Burkhardt2021,Johansen2021,Kooten2021}, as reservoirs of pebbles from different regions further outside the solar nebula could contribute to the volatile content in the forming carbonaceous chondrite parent bodies.

The chosen depletion factor of formaldehyde provided us with a conservative estimate of the least depletion predicted in the solar nebula models and thus indicated a possible upper limit for the initial formaldehyde concentrations in the source material of carbonaceous chondrite parent bodies. Other processes, e.g., the outgassing in these porous bodies heated by the decay of short-lived radioactive isotopes, was shown to potentially reduce their most volatile content even further \cite{Lichtenberg2021}.

TPD experiments with glycolaldehyde showed that its desorption was dictated by water ice \cite{Burke2014}. As we assumed water ice to stay frozen in the pebbles, we expected the glycolaldehyde abundance to be similar to cometary values and therefore we introduced no additional depletion factor. A more in-depth and detailed analysis of the depletion of volatiles was described in our previous study \cite{Paschek2021}.

It is important to note that our prediction of the concentrations of the volatile reactants was tailored to the view that carbonaceous planetesimals were assembled mostly instantaneously via streaming instabilities in the expected \SIrange{2}{3}{au} region inside the solar nebula. For this reason, it is strongly dependent on assumptions on the physico-chemical processes dominating the solar nebula and protoplanetary disc of the solar system.

Table~\ref{tab:init_concs} lists the concentrations of the considered reactants. The cometary concentrations were corrected by the depletion factor of formaldehyde in the correction factor column, giving rise to the solar nebula model-guided estimate for the initial concentrations in carbonaceous chondrite parent bodies in the predicted concentration column. The predicted concentrations were then used as the input parameters in our theoretical chemistry calculations for the abundances of ribose in this study.
 
\begin{table}[ht]
\setlength{\tabcolsep}{2.25mm}
\footnotesize
    \caption{Initial concentrations of reactants. The concentrations were normalized to water. The predicted concentrations using solar nebula models (except glycolaldehyde) were already used in the previous study \cite{Paschek2021}, which were the adjusted version of the ones found in comets \cite{Mumma2011} (and references therein) used by \citet{Cobb2015} and \citet{Pearce2016}. This correction was applied to be more representative of the concentrations present in pristine carbonaceous chondrite parent bodies. The predicted concentrations (the last column) are the ones used for the theoretical abundance calculations in this study and were derived by multiplying the cometary concentrations (the third column) with the correction factors (the fourth column). \label{tab:init_concs}}
    
\begin{tabular}{ccccc}
    \toprule
        \textbf{Molecule} & \multirow{2}{*}{\textbf{Name}} & \textbf{Cometary Concentration} & \multirow{2}{*}{\textbf{Correction Factor}} & \textbf{Predicted Concentration} \\
        \textit{\textbf{i}} & & {\boldmath{${[\mathrm{mol}_i\cdot{}\mathrm{mol}_{\ce{H2O}}^{-1}]}$}} & & {\boldmath{${[\mathrm{mol}_i\cdot{}\mathrm{mol}_{\ce{H2O}}^{-1}]}$}} \\
    \midrule
        \ce{H2O} & water & \phantom{(0.05--}1\phantom{.60)$\times$~$10^{-0}$} & - & \phantom{(0.05--}1\phantom{.60)$\times$~$10^{-0}$}\\
        \ce{CH2O} & formaldehyde & \phantom{(0.05--}6.60\phantom{)}$\times$~$10^{-4}$ & $10^{-3}$ & \phantom{(0.05--}6.60\phantom{)}$\times$~$10^{-7}$ \\
        \ce{C2H4O2} & glycolaldehyde & (0.05--4.00)~$\times$~$10^{-4}$ & - & (0.05--4.00)~$\times$~$10^{-4}$ \\
    \bottomrule
    \end{tabular}
\end{table}

\subsection{Computations}\label{sec:computations}

\textls[-15]{We calculated the amounts of ribose using the formose reaction pathway (see Equation~\eqref{equ:formose})} under the conditions inside the carbonaceous planetesimals. The thermochemical equilibrium calculations were performed using the software \textit{ChemApp}, distributed by GTT Technologies \cite{Petersen2007} (version 740 (6 April 2020), available online: \url{https://gtt-technologies.de/software/chemapp/}, accessed on 17 November 2021). The central input for these calculations with ChemApp was the Gibbs free energies of formation of the molecules involved in the modeled reaction. This Gibbs energy data were taken from the thermodynamic database \emph{CHNOSZ} \cite{Dick2019} (version 1.3.6 (16 March 2020), authored by Jeffrey M.\ Dick, available online: \url{https://www.chnosz.net}, accessed on 17 November 2021). Further, we used the \textit{GCC} compiler \cite{gcc} and the following software packages: \textit{pybind11} \cite{pybind11}, \textit{rpy2} \cite{rpy2}, \textit{NumPy} \cite{harris2020array}, \textit{SciPy} \cite{2020SciPy-NMeth}, and \textit{Matplotlib} \cite{matplotlib2007}.

\subsubsection{Planetesimal Model}

The environmental conditions, mainly the temperatures, inside the parent bodies were provided by the thermal planetesimal model developed by \citet{Lange2021}. This 1D model considered the radioactive decay of short- and long-lived isotopes as the heat source in the planetesimal interiors. It simulated the thermal evolution of planetesimals from their formation in the solar system until the present time. The available data provided us with the radial, time-dependent temperature profiles inside these bodies. The adjustable parameters for the simulation of these planetesimals were mainly their radius and the time of formation after calcium-aluminium-rich inclusions (CAI). As in our previous study \cite{Paschek2021}, we considered hypothetical planetesimals with radii of \SIrange{3}{150}{\kilo\meter} and times of formation after CAI of \SIrange{0.5}{3.5}{\mega\year}. A porosity of ${\phi = 0.2}$ was assumed in all models, representing a typical value found in carbonaceous chondrites \cite{Mason1963}.

One important aspect to note is that, compared to the more complex models of \citet{Lange2021}, the planetesimal models here are a simplified version adapted to parent bodies of carbonaceous chondrites.

\begin{figure}[p]
    \includegraphics[width=0.98\columnwidth]{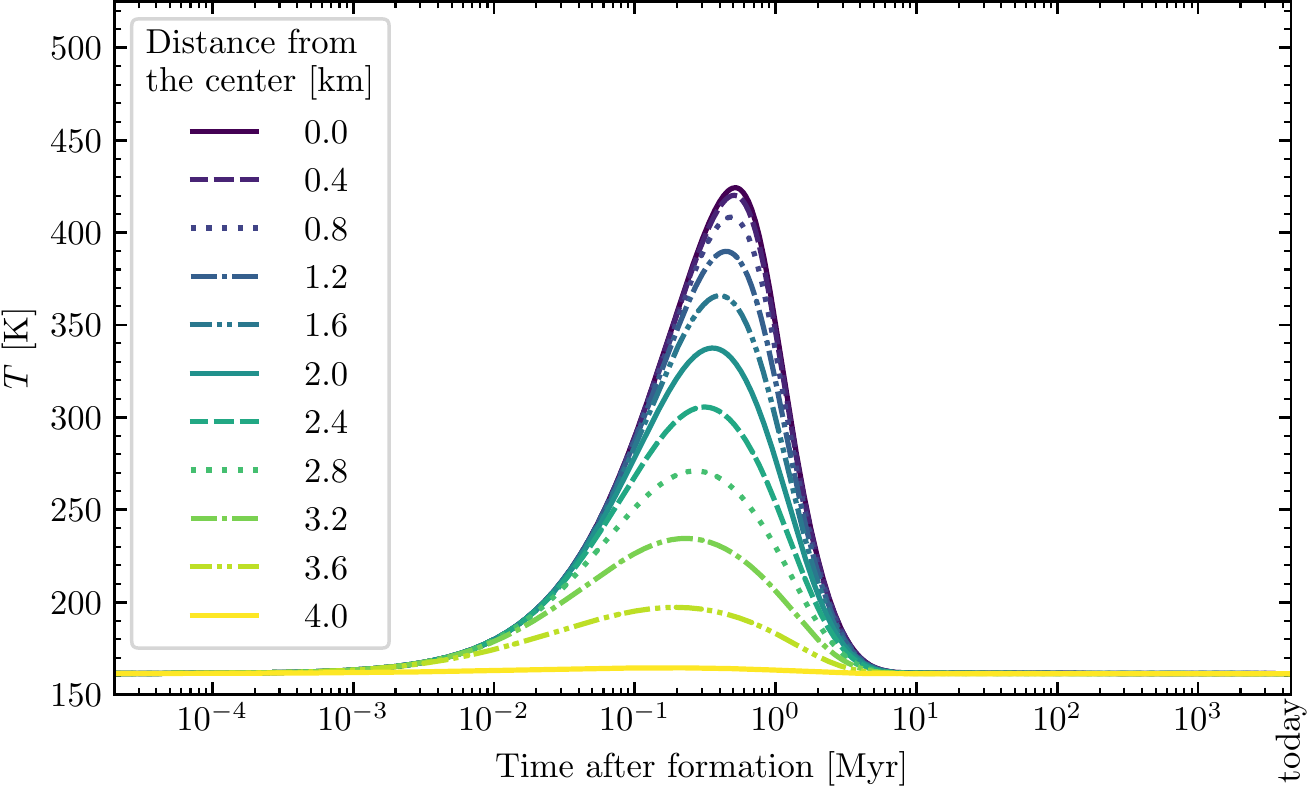}
    \caption{Temperature evolution inside a small- and early-formed model planetesimal over time. The temperature curves are given for different distances from the center inside the planetesimal. Properties of the planetesimal: ${\text{Radius} = \SI{4}{\kilo\meter}}$, porosity ${\phi = 0.2}$, and time of formation after calcium-aluminium-rich inclusions ${\text{(CAI)} = \SI{1}{\mega\year}}$. Reproduced from a simplified and adapted version of the model by \citet{Lange2021}. The temperature evolution for the other available model planetesimals was described in the previous study \cite{Paschek2021}.\label{fig:planetesimal_4km}}
\end{figure}

\begin{figure}[p]
    \includegraphics[width=0.98\columnwidth]{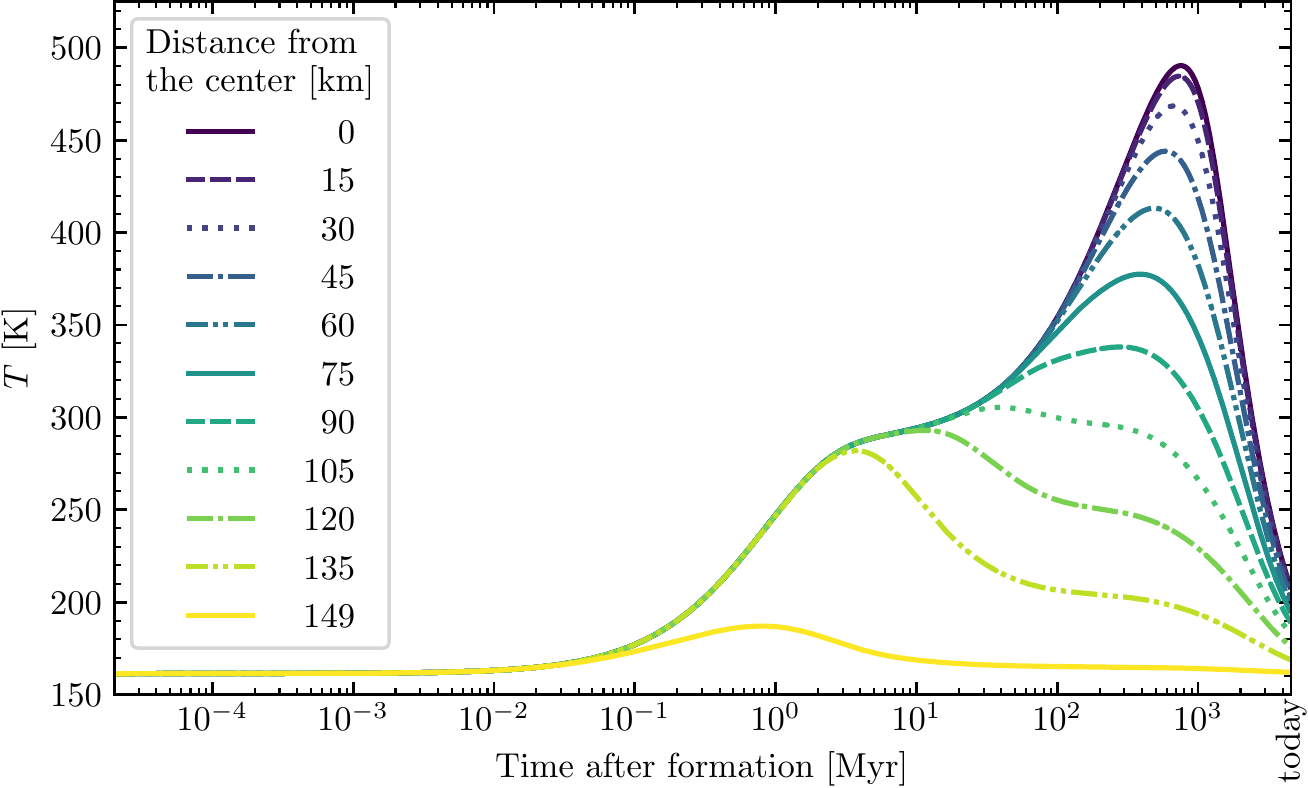}
    \caption{Temperature evolution inside a large- and late-formed model planetesimal over time. The temperature curves are given for different distances from the center inside the planetesimal. Properties of the planetesimal: ${\text{Radius} = \SI{150}{\kilo\meter}}$, porosity ${\phi = 0.2}$, and time of formation after ${\text{CAI} = \SI{3.5}{\mega\year}}$. Reproduced from a simplified and adapted version of the model by \citet{Lange2021}. The temperature evolution for the other available model planetesimals was described in the previous study \cite{Paschek2021}.\label{fig:planetesimal_150km}}
\end{figure}

Two examples of this planetesimal model can be seen in Figures~\ref{fig:planetesimal_4km}~and~\ref{fig:planetesimal_150km}. The first rise of temperature is caused by the decay of short-lived radionuclides (mainly \ce{^{26}Al}, with a small contribution of \ce{^{60}Fe}). For larger planetesimals, such as the one shown in Figure~\ref{fig:planetesimal_150km} with a radius of \SI{150}{\kilo\meter}, long-lived radionuclides (\ce{^{40}K}, \ce{^{232}Th}, \ce{^{235}U}, and \ce{^{238}U}) can also have a significant contribution. This results in a second temperature rise or plateau in the outer shells of the planetesimal over time. 

For a more comprehensive overview of the software, the Gibbs free energies of formation, and more details about the planetesimal model, we refer to our previous study~\cite{Paschek2021}. The pressure dependence of the Gibbs free energies of formation is very marginal, allowing us to assume \SI{100}{\bar} for all thermochemical calculations inside the entire planetesimal interiors. Further information about the thermochemical equilibrium calculations can be found in \citet{Pearce2016} and \citet{Cobb2015}. The source code, excluding the proprietary ChemApp library, and including the data of the planetesimal models, is openly available on Zenodo at (\cite{klaus_paschek_2021_5774880}, \url{https://doi.org/10.5281/zenodo.5774880}, accessed on 1 March 2022) and as a Git repository: \url{https://github.com/klauspaschek/prebiotic_synthesis_planetesimal}, accessed on 17 November 2021.

\subsubsection{Gibbs Free Energies of Formation of Glycolaldehyde}\label{sec:gibbs_glycolaldehyde}

\begin{figure}[t]
    \includegraphics[width=\textwidth]{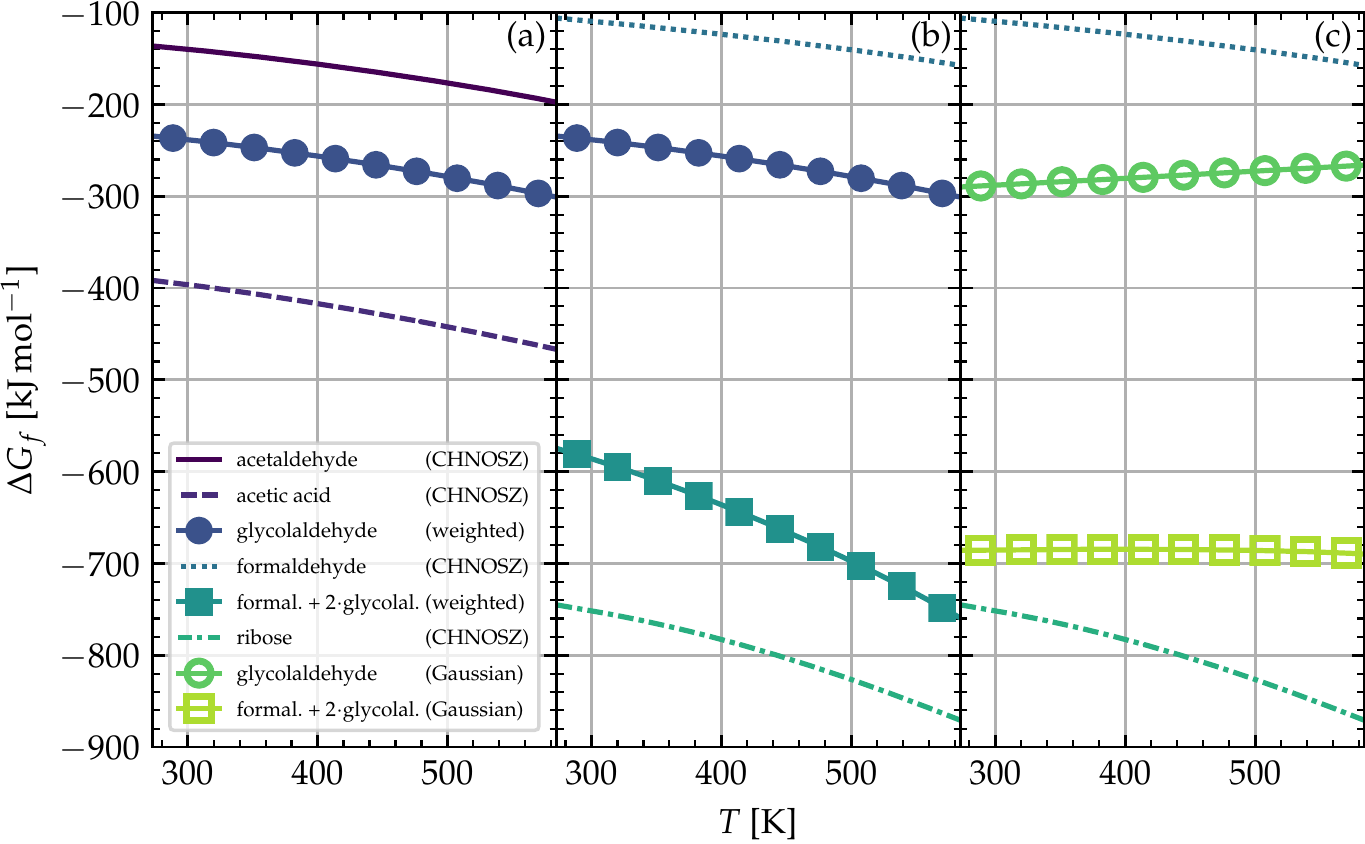}
    \caption{Gibbs free energies of formation of different molecules plotted against temperature at \SI{100}{\bar}. The energies for glycolaldehyde were estimated values calculated (\textbf{a},\textbf{b}) either by the technique developed by \citet{Emberson2010} and \citet{Fernandez2010}, and used in \citet{Cobb2015} (denoted in figure legend as ``weighted'', plotted as solid lines with filled symbols), (\textbf{c}) or by using the computational quantum chemistry software Gaussian \cite{g09} (denoted in figure legend as ``Gaussian'', plotted as solid lines with hollow symbols). Data taken from the CHNOSZ database are plotted as solid and dashed lines without symbols.\label{fig:gibbs_glycolal}}
\end{figure}

The CHNOSZ database used in our modeling does not contain the Gibbs energies for glycolaldehyde. Therefore, we provide and compare the two ways of estimating these energies, which we used in our calculations. \citet{Cobb2015} gave an estimate by modeling glycolaldehyde as a mixture of acetaldehyde and acetic acid, for which the CHNOSZ database does have the Gibbs energies. The motivation is that the combination of acetaldehyde and acetic acid roughly resembles glycolaldehyde's functional groups and structure. In the works of \citet{Emberson2010} and \citet{Fernandez2010}, the respective enthalpies of formation $\Delta H_f$ for acetaldehyde and acetic acid were weighted to fit the one of glycolaldehyde given by \citet{Espinosa-Garcia2005} at standard conditions. The weights were found to be \SI{61.1}{\percent} and \SI{38.2}{\percent}, respectively. The same weighting coefficients were used to weight the Gibbs free energies of formation $\Delta G_f$ of acetaldehyde and acetic acid (taken from CHNOSZ) to estimate the one for glycolaldehyde. These weighted energies can be found in Figure~\ref{fig:gibbs_glycolal}a.

For the second method of estimation, we performed quantum chemistry calculations using the \textit{Gaussian 09} software package \cite{g09} (available online: \url{https://gaussian.com/}, accessed on 17 November 2021). These quantum chemistry calculations were used to obtain the atomic and molecular energies and entropies necessary to directly calculate the Gibbs free energy of formation for glycolaldehyde. All quantum chemistry calculations were performed using the \text{Becke-3–Lee–Yang–Parr} (B3LYP) hybrid density functional~\cite{Stephens1994,Becke1993,Lee1988} and the polarizable continuum model (PCM) for aqueous solution effects \cite{Miertus1981,Cammi1995}. Geometries were optimized using the \text{6-31G(d,p)} basis set, and single-point energies and frequencies were calculated at a higher level basis set, i.e., \text{6-311++G(2df,2p)}, using the geometries optimized in the previous step. This particular method was used in the past by \citet{Espinosa-Garcia2005} to calculate the enthalpy of glycolaldehyde (see first estimation method above). All calculations were performed at \SI{100}{\bar} to match the peak pressures inside the meteoritic parent bodies \cite{Pearce2016}, for reasons explained in previous studies (see Section~\ref{sec:computations} above). The Gibbs free energies of formation were calculated using the three-step method outlined by \citet{Ochterski2000}, i.e., (1) calculate the enthalpy of formation at \SI{0}{\kelvin}, (2) calculate the enthalpy of formation at \SI{298}{\kelvin} from elements in their standard states, and (3) calculate the entropy of formation $\Delta S_f$ from elements in their standard states at \SI{298}{\kelvin}, and insert everything into the standard Gibbs formula ${\Delta G_f = \Delta H_f - T\Delta S_f}$.

The atomic and molecular energies and entropies used for the $\Delta G_f$ calculations were obtained from our quantum chemistry calculations. The only exceptions were the heats of formation for the atoms and the entropy for standard state carbon (graphite) at \SI{298}{\kelvin}, which were obtained from experiments \cite{Curtiss1997} and National Bureau of Standards (current name: National Institute of Standards and Technology) tables of the chemical thermodynamic properties \cite{Wagman1982}, respectively. Due to the lack of experimental entropic data for carbon (graphite) above \SI{298}{\kelvin}, a \SI{1}{\percent} increase per \SI{25}{\kelvin} was introduced to the experimental carbon (graphite) entropy. This correction was done to match similar entropy increases from our quantum chemistry calculations of hydrogen and oxygen. Lastly, the atomic enthalpy correction was calculated regarding the gas-state carbon rather than carbon (graphite), which introduced ${\sim}$$\SI{3}{\kilo\joule\per\mole}$ error into our calculations.

We validated this method by calculating $\Delta G_f$ for acetaldehyde and formaldehyde at \SI{298}{\kelvin}. Our calculated values were within \SI{6}{\kilo\joule\per\mole} and \SI{10}{\kilo\joule\per\mole} of the values from the CHNOSZ database, respectively. Improvements may be made to this method by calculating enthalpies and entropies for carbon (graphite) to be consistent with the quantum calculations of the other entropies. However, given that the results of the ribose synthesis in this study were not sensitive to our glycolaldehyde calculations for neither the first nor this second estimate (see the explanation below), we did not further improve upon these calculations this time. The resulting second estimate for $\Delta G_f$ of glycolaldehyde is plotted in Figure~\ref{fig:gibbs_glycolal}c.

Although the two estimates for $\Delta G_f$ of glycolaldehyde were slightly different, they both showed the same theoretical results for the ribose abundances. This was because the adopted reaction pathway for the synthesis of ribose is limited by the initial abundances of reactants, and hence is less sensitive to the $\Delta G_f$ values. This was verified by analyzing the output abundances from ChemApp, which showed that all initially present reactants had zero abundances after ribose was formed. 

Consequently, comparing the Gibbs energies of reactants and products allows us to determine whether ribose should be formed. When looking at the modeled reaction pathway for ribose (see Equation~\eqref{equ:formose}), one molecule of formaldehyde and two of glycolaldehyde are combined to form one molecule of ribose. Following this stoichiometry, one has to compare the Gibbs energies of ribose with the sum of the energies of formaldehyde and twice the energies of glycolaldehyde (\ce{1\text{formaldehyde} + 2\text{glycolaldehyde} -> 1\text{ribose}}, Equation~\eqref{equ:formose}). Ribose should only be formed if the energies for ribose are more negative than the stoichiometric sum of those for formaldehyde and glycolaldehyde.

Figure~\ref{fig:gibbs_glycolal}b shows this sum compared to ribose with the first weighted estimate for glycolaldehyde, and Figure~\ref{fig:gibbs_glycolal}c shows the comparison with the second estimate with Gaussian. The stoichiometric sum of the Gibbs energies of formaldehyde and glycolaldehyde was less negative than the Gibbs energies of ribose in both cases. Thus, indeed, only the limitation by the initial abundances of reactants mattered for the ribose formation. If the stoichiometric sum of formaldehyde and glycolaldehyde crossed with the energies of ribose, no ribose would be formed at the respective temperatures. Therefore, both estimates of the Gibbs energies of glycolaldehyde allowed us to model this reaction pathway in the same way.

\subsection{Laboratory Experiments}\label{sec:lab}

There are about 40 different products formed in the formose reaction \cite{Rauchfuss2005}. These include plenty of sugars such as aldoses and ketoses, sugar alcohols, sugar acids, branched sugars, and even decomposition products such as lactic acid, see also, e.g., \cite{Omran2020}. Modeling the complex formose network with its different reactions such as aldol reactions, retro-aldol reactions, arrangements, and decomposition reactions is a big challenge for theoretical and analytical chemistry, see, e.g., \cite{Kim2011}. Accordingly, following the kinetics of every single molecule is almost impossible, and the thermodynamic data of the molecules and reactions are often  missing.

Therefore, the reaction pathway considered here and summarized in Equation~\eqref{equ:formose} is a major simplification. To compensate for this, we performed laboratory experiments of the formose reaction in an aqueous solution and measured the amounts of the resulting sugars. This allowed us to find the fraction of ribose in all forming 5Cs.

In our experiments, the reaction was started with a very concentrated solution of formaldehyde (\SI{1.34}{\mole\per\liter}) and glycolaldehyde (\SI{0.269}{\mole\per\liter}, \SI{20}{\mole\percent}) based on the prebiotic DNA formation pathway demonstrated by \citet{Teichert2019} (for more context, see the review by \citet{Kruse2020}). To this solution, \SI{10}{\mole\percent} of one of the catalysts were added, and the temperature was kept stable. At different time intervals, depending on the activity of the catalyst, a sample was taken. The whole solution had a volume of \SI{1}{\milli\liter} at the start of each run, and samples were taken in volumes of \SI{50}{\micro\liter} each. The formose reaction was stopped by adding citric acid and freezing the sample. After lyophilization, the sample was analyzed via gas chromatography in combination with a mass spectrometer, as described in the preceding studies by \citet{Haas2018,Haas2020} (abbreviated there as GC-MS).

There are also other possibilities to analyze products in the formose reaction, e.g., coupling liquid chromatography with UV and electrospray ionization-mass spectrometry (abbreviated as LC-UV and ESI-MS) \cite{Zweckmair2013}. This other method could help overcome problems, e.g., thermal instability of some of the analyzed molecules in GC-MS analysis, and could be interesting for follow-up studies.

The maximum yields of ribose in all 5Cs inferred from our measurements were used as a correction factor for the ribose abundances resulting from our theoretical thermochemical equilibrium model.

\section{Results}\label{sec:results}

\subsection{Experimentally Found Yields of Ribose in All 5Cs}\label{sec:yields}

\begin{figure}[t]
    \includegraphics[width=\textwidth]{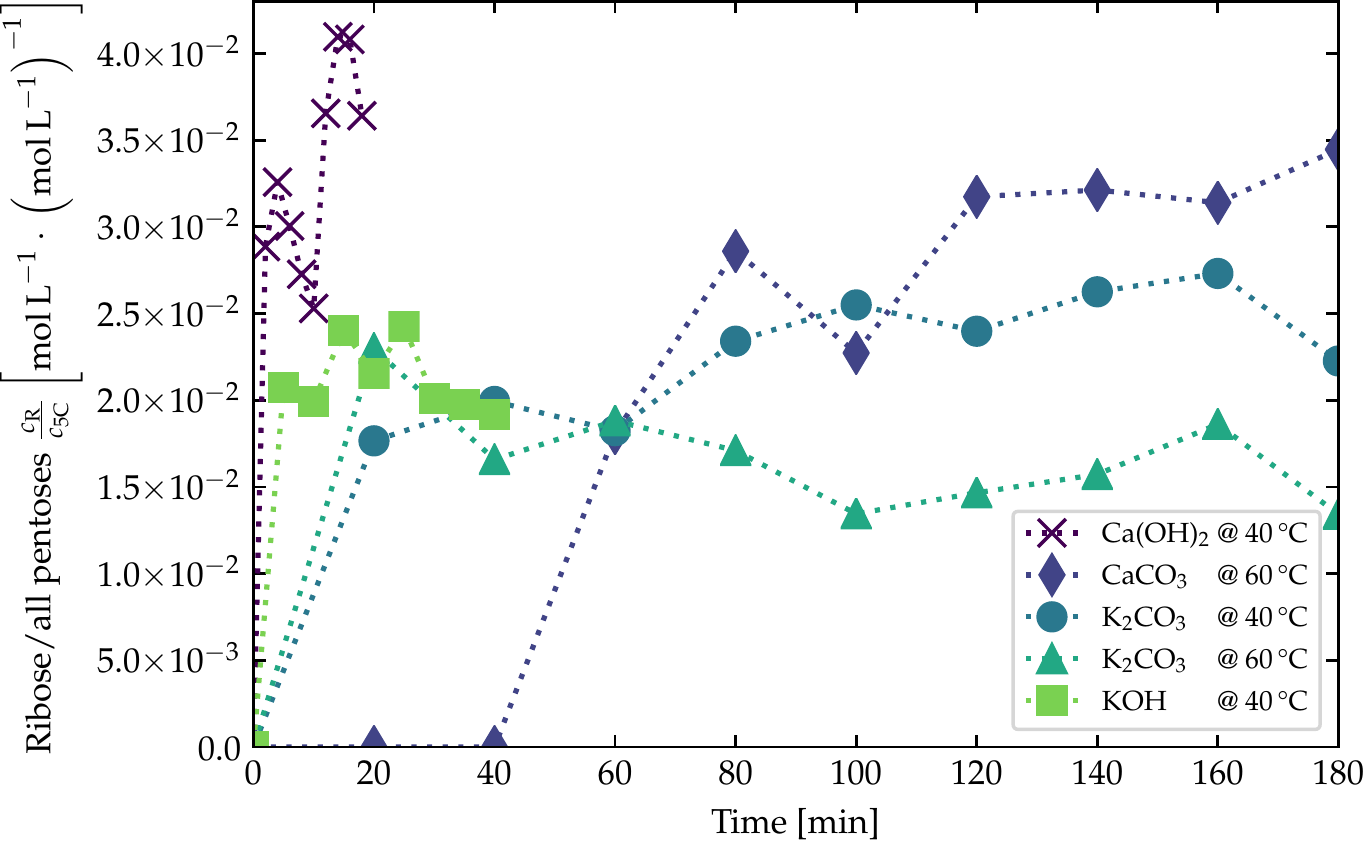}
    \caption{Fraction of ribose (R) in all pentoses (5Cs) synthesized over time in lab experiments. The reaction started with concentrations for formaldehyde of \SI{1.34}{\mole\per\liter}, for glycolaldehyde of \SI{0.269}{\mol\per\liter} or \SI{20}{\mole\percent}, and for the respective catalyst of \SI{10}{\mole\percent}, in a solution volume of \SI{1}{\milli\liter}, with samples taken over time in volumes of \SI{50}{\micro\liter} each. The temperature of the solution was kept constant over time at the values denoted in the figure legend for each catalyst run.\label{fig:yields}}
\end{figure}

Figure~\ref{fig:yields} shows our experimental results for the fraction of ribose in all 5Cs plotted against the elapsed time of the reaction. Calcium hydroxide was the catalyst with the highest activity. Compared to the other catalysts, it formed the highest amount of ribose in the shortest amount of time. However, one has to be aware that the decomposition of the formose products is also more rapid (see Discussion in Section~\ref{sec:discussion}). Focusing on ribose, it took less than \SI{20}{\minute} to reach its maximum yield. Likewise, potassium hydroxide and potassium carbonate produced ribose very quickly in significant amounts, with the maximum values being lower than those obtained with calcium hydroxide. Calcium carbonate took significantly longer to produce higher sugars, which can be explained by its lower solubility. Since we used \SI{10}{\mole\percent} of the catalyst in each experiment, calcium carbonate reached its maximum aqueous solubility. This was done in order to stay consistent within the framework of our experiments and with the previous study by \citet{Teichert2019}. Still, after \SI{180}{\minute} the experimental run with calcium carbonate resulted in the second-highest yield for ribose. All catalysts did not seem to differ much in effectiveness. They produced ribose with yields of the same order of magnitude.

The maximum yields of ribose reached in Figure~\ref{fig:yields} are listed in Table~\ref{tab:yields}. These values were the correction factors used in the theoretical model to compensate for the oversimplification of the modeled reaction in Equation~\eqref{equ:formose}.

\vspace{-6pt}

\begin{table}[ht]
\setlength{\tabcolsep}{7.65mm}
\small
    \caption{Maximum yield of ribose (R) in all 5Cs reached in lab experiments for different present catalysts, respectively (see Figure~\ref{fig:yields}). The reaction always started from formaldehyde and glycolaldehyde as reactants in aqueous solution.\label{tab:yields}}
    \begin{tabular}{ccc}
    \toprule
    \multirow{2}{*}{\textbf{Catalyst}\vspace{-6pt}} & \multirow{2}{*}{\textbf{Name}\vspace{-6pt}} & \textbf{Maximum Yield of Ribose} \\
    & & \textbf{{\boldmath{$\left\{\frac{c_{\text{R}}}{c_{\text{5C}}}\right\}_{\mathrm{max}}\,{\left[{\mathrm{mol~L^{-1}}\cdot{}\left(\mathrm{mol~L^{-1}}\right)^{-1}}\right]}$}}}\\
    \midrule
    \ce{Ca(OH)2} & Calcium hydroxide & 4.1 $\times$ $10^{-2}$ \\
    \ce{CaCO3} & Calcium carbonate & 3.5 $\times$ $10^{-2}$ \\
    \ce{K2CO3} & Potassium carbonate & 2.7 $\times$ $10^{-2}$ \\
    \ce{KOH} & Potassium hydroxide & 2.4 $\times$ $10^{-2}$ \\
    \bottomrule
    \end{tabular}
\end{table}

\newpage
\subsection{Theoretically Calculated Ribose Abundances in Planetesimals}\label{sec:ribose}
\begin{figure}[t]
    \includegraphics[width=\textwidth]{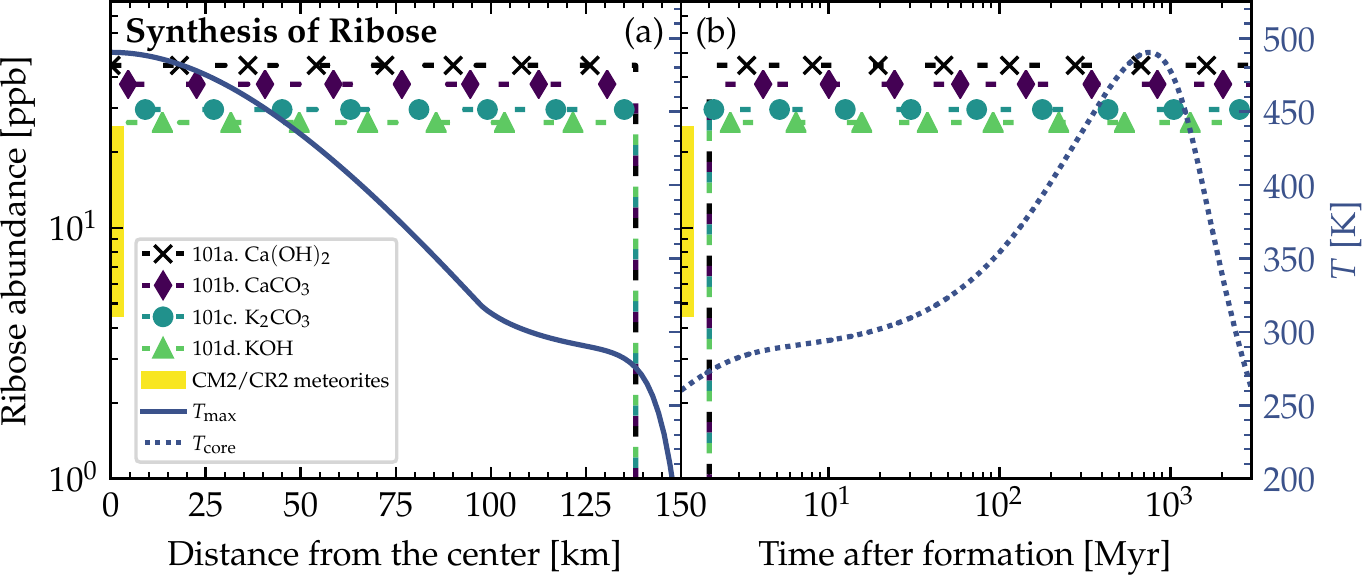}
    \caption{Lower bound theoretical ribose abundances from simulations of formose reaction pathway in Equation~\eqref{equ:formose}. Properties of planetesimal: ${\text{Radius} = \SI{150}{\kilo\meter}}$, densities ${\rho_{\mathrm{rock}} = \SI{3}{\gram\per\centi\meter\cubed}}$, ${\rho_{\mathrm{ice}} = \SI{0.917}{\gram\per\centi\meter\cubed}}$, porosity ${\phi = 0.2}$, and time of formation after ${\text{CAI} = \SI{3.5}{\mega\year}}$. The experimentally found yields of ribose within 5Cs for each catalyst (see Table~\ref{tab:yields}) were multiplied with the theoretically calculated 5C abundance to obtain the ribose abundances (dashed lines with symbols). This simulation was run with the \textit{lower} (opposite to Figure~\ref{fig:R_upper}) bound of the initial concentration of glycolaldehyde of ${5\times10^{-6}\,\mathrm{mol}\cdot{}\mathrm{mol}_{\ce{H2O}}^{-1}}$ (see Table~\ref{tab:init_concs}). All simulations were run at \SI{100}{bar}. In both panels (\textbf{a}) and (\textbf{b}) the left vertical axis corresponds to the abundances (dashed lines with symbols) and the right vertical axis corresponds to the temperatures from the planetesimal model (solid and dotted lines). The shaded part of the abundance axis represents the range of ribose abundances measured in CM2 (Mighei-type, Murchison, upper limit) and CR2 (Renazzo-type, NWA 801, lower limit) meteorites \cite{Furukawa2019}, and has no correlation to the radial location inside the object or the point in time (horizontal axes). (\textbf{a}) Distribution of abundances for the maximum temperature $T_{\mathrm{max}}$ (solid line) reached at a specific distance from the center inside the planetesimal (center at the left and surface at the right). Ribose was synthesized at and below \SI{138}{\kilo\meter} distance from the center. (\textbf{b}) Evolution of abundances at temperatures $T_{\mathrm{core}}$ (dotted line) in the center of the planetesimal over time (the same temperature evolution curve can be found in Figure~\ref{fig:planetesimal_150km}). Ribose synthesis started at \SI{2}{\mega\year} after formation.\label{fig:R_lower}}
\end{figure}

\begin{figure}[t]
    \includegraphics[width=\textwidth]{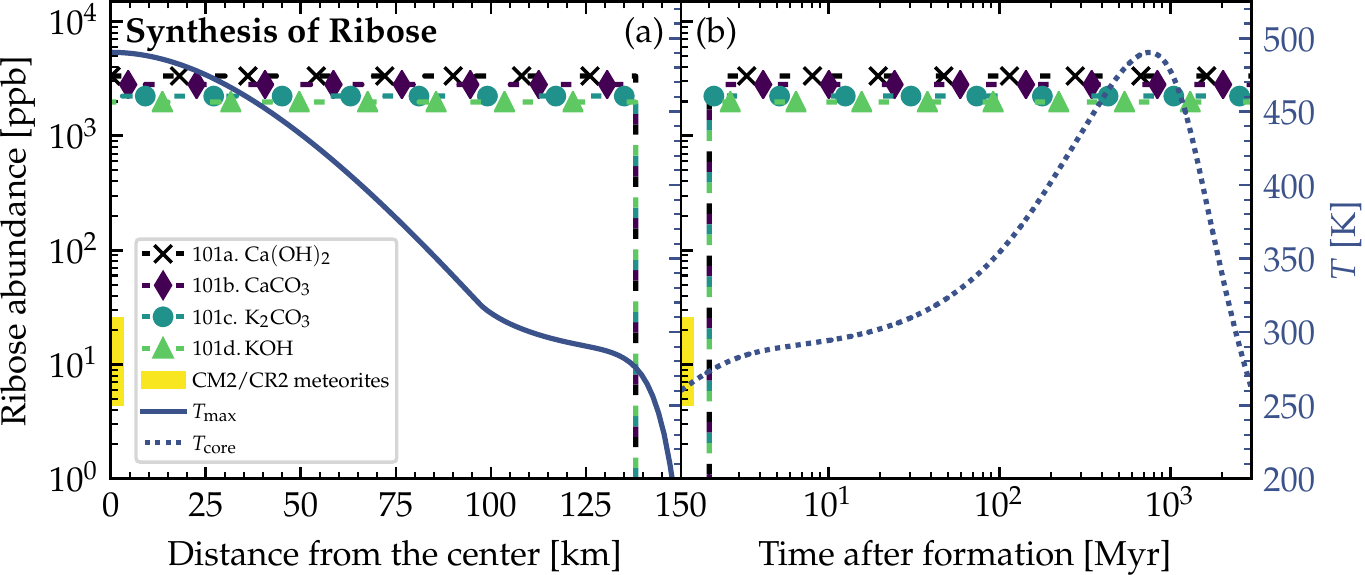}
    \caption{Upper bound theoretical ribose abundances from simulations of formose reaction pathway in Equation~\eqref{equ:formose}. Properties of planetesimal: ${\text{Radius} = \SI{150}{\kilo\meter}}$, densities ${\rho_{\mathrm{rock}} = \SI{3}{\gram\per\centi\meter\cubed}}$, ${\rho_{\mathrm{ice}} = \SI{0.917}{\gram\per\centi\meter\cubed}}$, porosity ${\phi = 0.2}$, and time of formation after ${\text{CAI} = \SI{3.5}{\mega\year}}$. The experimentally found yields of ribose within 5Cs for each catalyst (see Table~\ref{tab:yields}) were multiplied with the theoretically calculated 5C abundance to obtain the ribose abundances (dashed lines with symbols). This simulation was run with the \textit{upper} (opposite to Figure~\ref{fig:R_lower}) bound of the initial concentration of glycolaldehyde of ${4\times10^{-4}\,\mathrm{mol}\cdot{}\mathrm{mol}_{\ce{H2O}}^{-1}}$ (see Table~\ref{tab:init_concs}). All simulations were run at \SI{100}{bar}. In both panels (\textbf{a}) and (\textbf{b}) the left vertical axis corresponds to the abundances (dashed lines with symbols) and the right vertical axis corresponds to the temperatures from the planetesimal model (solid and dotted lines). The shaded part of the abundance axis represents the range of ribose abundances measured in CM2 (Murchison, upper limit) and CR2 (NWA 801, lower limit) meteorites \cite{Furukawa2019}, and has no correlation to the radial location inside the object or the point in time (horizontal axes). (\textbf{a}) Distribution of abundances for the maximum temperature $T_{\mathrm{max}}$ (solid line) reached at a specific distance from the center inside the planetesimal (center at the left and surface at the right). Ribose was synthesized at and below \SI{138}{\kilo\meter} distance from the center. (\textbf{b}) Evolution of abundances at temperatures $T_{\mathrm{core}}$ (dotted line) in the center of the planetesimal over time (the same temperature evolution curve can be found in Figure~\ref{fig:planetesimal_150km}). Ribose synthesis started at \SI{2}{\mega\year} after formation.\label{fig:R_upper}}
\end{figure}

The resulting ribose abundances were calculated for the temperatures taken from a simplified version of the planetesimal model by \citet{Lange2021} (see Section~\ref{sec:computations}). The temperatures inside the planetesimal are plotted as solid and dashed lines in Figures~\ref{fig:R_lower}~and~\ref{fig:R_upper} and correspond to the example planetesimal model shown in Figure~\ref{fig:planetesimal_150km}. Figures~\ref{fig:R_lower}~and~\ref{fig:R_upper} show the resulting ribose abundances for several parameters and are a representative selection among all the available thermal profiles. We also calculated the abundances for the other available planetesimal models with our chemical simulations, but the results share the same characteristics and trends as for the shown planetesimal model in Figures~\ref{fig:R_lower}~and~\ref{fig:R_upper} (for the other available planetesimal models see also the previous study \cite{Paschek2021}). The resulting ribose abundances for different catalyst correction factors (see Table~\ref{tab:yields}) are plotted as dashed lines with symbols (see legends). The shaded part of the abundance axis in both figures represents the range of measured ribose concentrations in carbonaceous chondrites~\cite{Furukawa2019}. These measured abundances represent the to-be-achieved benchmark and real-world reference to our theoretically calculated values.

Figures~\ref{fig:R_lower}a~and~\ref{fig:R_upper}a show the radial distribution of the calculated ribose abundances in the planetesimal's interior. The maximum temperature $T_{\mathrm{max}}$ (solid line) reached at a specific distance from the center inside the planetesimal over the entire simulation time (from the planetesimal's formation until today) was taken into account to calculate the ribose abundances. In this case, we inherently assumed that the peak production of the ribose sugar was achieved at the peak temperature at each distance from the center. The left side of these panels (a) defines the center, and the right side the surface of the planetesimal.

In Figures~\ref{fig:R_lower}b~and~\ref{fig:R_upper}b, the dashed curve shows the temperature $T_{\mathrm{core}}$ in the center of the planetesimal over time. The ribose abundances were calculated by iterating over time and using the abundances of reactants and products resulting from previous steps as the initial abundances in each step. This allowed us to follow the equilibrium of the reaction proceeding through time.

In Figure~\ref{fig:R_lower}, the \textit{lower} bound of the initial concentration of glycolaldehyde of $5\times10^{-6}\,\mathrm{mol}\cdot{}\mathrm{mol}_{\ce{H2O}}^{-1}$ from Table~\ref{tab:init_concs} was used. As a result, the simulated ribose abundances coincided with measurements in carbonaceous chondrites \citep{Furukawa2019} within a factor of \numrange{2}{3}. On the other hand, in Figure~\ref{fig:R_upper} we used the \textit{upper} bound of $4\times10^{-4}\,\mathrm{mol}\cdot{}\mathrm{mol}_{\ce{H2O}}^{-1}$, and the calculated ribose abundances were higher than the measured values by about two orders of magnitude.

As the reduction of the initial cometary formaldehyde concentration by $10^{-3}$ (see Table~\ref{tab:init_concs}) was a very broad assumption, a change of this reduction will change the resulting ribose abundances. Keeping this in mind, the theoretically simulated abundances can be considered to be in reasonable agreement with the measured ones. Further, we suspect that decomposition and other complex reactions that we did not take into account could be responsible for the slightly overestimated theoretical ribose abundances.

\vspace{-6pt}

\section{Discussion and Conclusions}\label{sec:discussion}

In the thermochemical equilibrium calculations, the resulting abundances of the reaction products depend strongly on the initial concentrations of the reactants (see Table~\ref{tab:init_concs}). Therefore, the resulting ribose abundances in Figures~\ref{fig:R_lower}~and~\ref{fig:R_upper} were strongly dependent on the initial abundance of glycolaldehyde. Furthermore, a different estimate for the initial abundance of formaldehyde and the proper rescaling of the pristine cometary values guided by solar nebula models would also change the resulting ribose concentrations. This rescaling had to be made and is hard to verify via measurements, observations, or modeling, since the icy pebbles out of which the carbonaceous chondrite planetesimals formed in the solar nebula have been gone for too long. Considering all these limitations, uncertainties, and the fact that our correction factor for the initial formaldehyde abundance was an upper limit, the simulated and measured ribose abundances shown in Figures~\ref{fig:R_lower}~and~\ref{fig:R_upper} still coincide reasonably well.

We see this as the confirmation that the formose reaction could be the pathway to forming sugars such as ribose abiotically. Pristine carbonaceous chondrites are time capsules showing us how the foundations for the emergence of life might have been laid in our early solar system.

Ribose is susceptible to decomposition, see, e.g., \cite{Larralde1995}. Particularly at higher temperatures further away from the freezing point of water, significant portions of ribose and other sugars could be destroyed or converted to even more complex species (e.g., polysaccharides). In this context, typical decomposition processes of sugars are the $\beta$-elimination to dicarbonyls, the benzilic acid rearrangement, oxidation, and others, see, e.g., \cite{DeBruijn1986}. In laboratory experiments, the solution containing the freshly formed sugars starts to turn yellow (the so-called ``yellowing point'') when the maximum abundances of sugars are reached. This corresponds roughly to the point in time when the maximum fraction of ribose was reached in Figure~\ref{fig:yields}. After a while, the solution starts to turn brownish (forming so-called ``brown tar'') as the decomposition proceeds.

Since we used the maximum yields of ribose reached in the experiments (see Table~\ref{tab:yields}) as the correction for our theoretical studies, our results represent an estimate of the possible upper limit. This probably led to the slightly too high ribose abundances when compared to the measurements in carbonaceous chondrites (see Figures~\ref{fig:R_lower}~and~\ref{fig:R_upper}). Decomposition plus other reactions could have lowered the ribose abundance in the planetesimals over time, which might explain the lower values measured in carbonaceous chondrites.

Our results seemed to be reasonable when taking these potentially adverse effects into account. Ribose is still found today in meteorites \cite{Furukawa2019}, indicating that it only decomposed to a certain extent. \citet{Ricardo2004} showed in laboratory experiments that boron from borate minerals stabilized ribose and other 5Cs in their cyclic furanose form. The solution did not turn brownish for 2 months as decomposition was prevented. They also postulated that boron could stabilize glyceraldehyde, an intermediate reaction product in the formose reaction, keeping it from decomposing into ``brown tar'', and therefore enhancing the formation of complex sugars from glycolaldehyde and glyceraldehyde. Since boron was found in carbonaceous chondrites \cite{Mills1968}, it could stabilize the formed ribose in meteorites and their parent bodies.

Ribose destruction rates by hydrolysis could be dampened by many orders of magnitude at temperatures below \SI{60}{\celsius} at pH between \numrange{4}{8} \cite{Larralde1995}. The outer shells of planetesimals were heated only for short periods of time to temperatures above the melting point of water before they were frozen again. Therefore, in these outer shells, the freshly formed ribose (besides other prebiotic molecules) might be quickly frozen in the water. This potentially reduced the chance of decomposition as the temperatures were lower compared to the core region of the planetesimal, and liquid water only existed over a shorter period of time. The frozen ribose might have been preserved until it was distributed in the solar system as fragments of the parent body, and some of the fragments fell to the Earth as meteorites. Extraterrestrial organic and prebiotic molecules were found in meteorites on the Earth's surface today, see, e.g., \cite{Furukawa2019,Cobb2014,Pearce2015,Lai2019,Gilmour2003,Pizzarello2006,Derenne2010}, although the possibility of at least partial terrestrial contamination has to be considered. Theoretical studies coupled with experimental data~\cite{Chyba1990,Chyba:1992bp,Basiuk1998,Pierazzo1999} suggested that a significant portion of organics might survive the heating due to friction in the atmosphere and the energy of the impact \cite{Brinton1996} and arrive intact on the Earth's surface even in comets and interplanetary dust particles. Destruction during the atmospheric entry and impact could be another reason why the detected abundances of ribose in carbonaceous chondrites were lower than our calculated results.

\begin{figure}[p]
    \includegraphics[width=0.82\columnwidth]{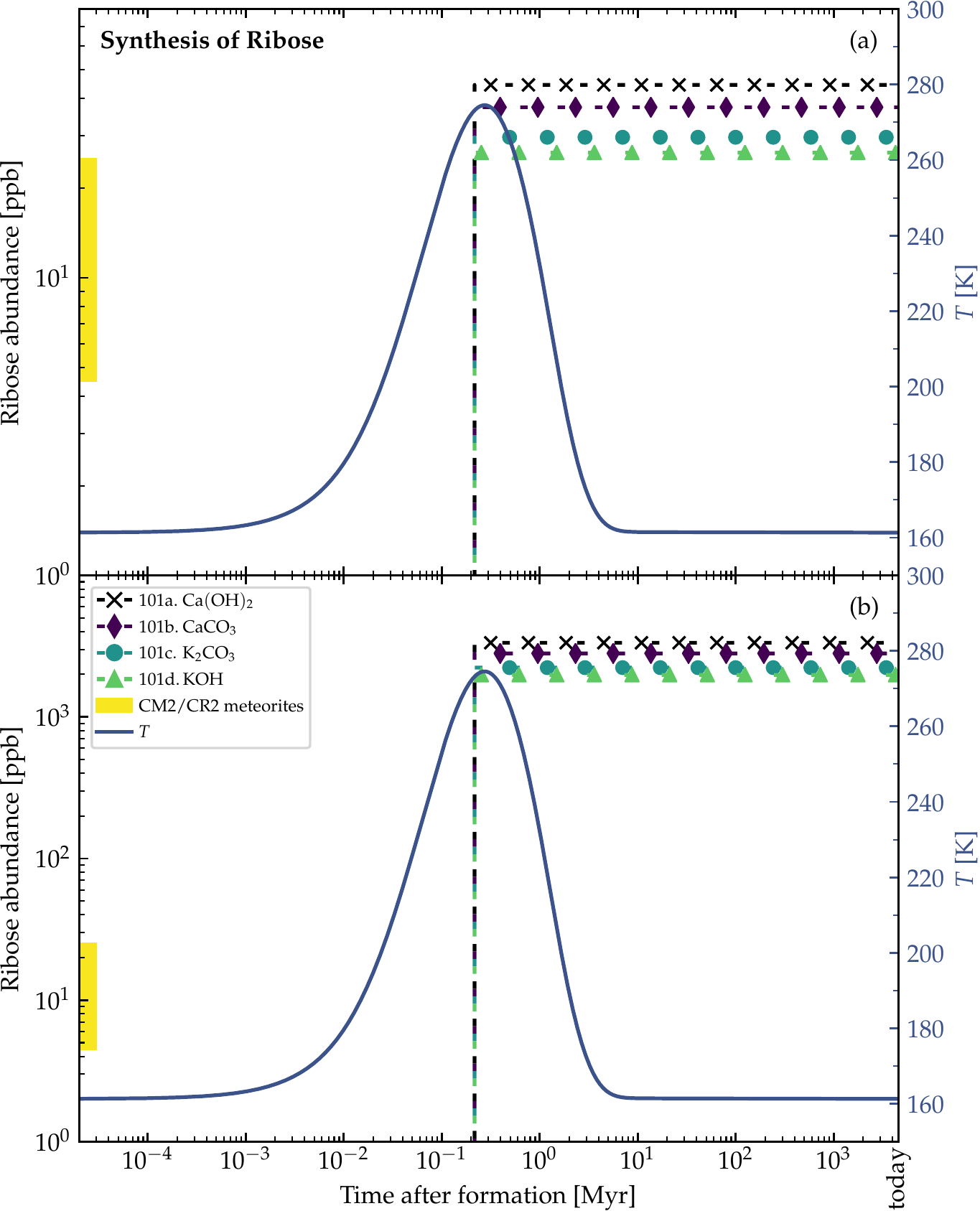}
    \caption{Theoretical ribose abundances in an outer shell at \SI{2.76}{\kilo\meter} distance from the center of the \SI{4}{\kilo\meter}-sized planetesimal model. The whole planetesimal model is shown in Figure~\ref{fig:planetesimal_4km}. Properties of planetesimal: ${\text{Radius} = \SI{4}{\kilo\meter}}$, densities ${\rho_{\mathrm{rock}} = \SI{3}{\gram\per\centi\meter\cubed}}$, ${\rho_{\mathrm{ice}} = \SI{0.917}{\gram\per\centi\meter\cubed}}$, porosity ${\phi = 0.2}$, and time of formation after ${\text{CAI} = \SI{1}{\mega\year}}$. The formose reaction pathway in Equation~\eqref{equ:formose} was used in the simulations. The experimentally found yields of ribose within 5Cs for each catalyst (see Table~\ref{tab:yields}) were multiplied with the theoretically calculated 5C abundance to obtain the ribose abundances (dashed lines with symbols). Ribose synthesis started at ${\sim}$210,000 yr after formation. All simulations were run at \SI{100}{bar}. In both panels (\textbf{a},\textbf{b}), the left vertical axis corresponds to the abundances (dashed lines with symbols) and the right vertical axis corresponds to the temperatures $T$ (solid lines) in the outer shell of the planetesimal model. The shaded part of the abundance axis represents the range of ribose abundances measured in CM2 (Murchison, upper limit) and CR2 (NWA 801, lower limit) meteorites \cite{Furukawa2019}, and has no correlation to the point in time (horizontal axis). (\textbf{a}) Time evolution of lower bound abundances simulated using the \textit{lower} (opposite to panel (\textbf{b})) bound of the initial concentration of glycolaldehyde of ${5\times10^{-6}\,\mathrm{mol}\cdot{}\mathrm{mol}_{\ce{H2O}}^{-1}}$ (see Table~\ref{tab:init_concs}). (\textbf{b}) Time evolution of upper bound abundances simulated using the \textit{upper} (opposite to panel (\textbf{a})) bound of the initial concentration of glycolaldehyde of ${4\times10^{-4}\,\mathrm{mol}\cdot{}\mathrm{mol}_{\ce{H2O}}^{-1}}$.\label{fig:shell_4km}}
\end{figure}

\begin{figure}[p]
    \includegraphics[width=0.82\columnwidth]{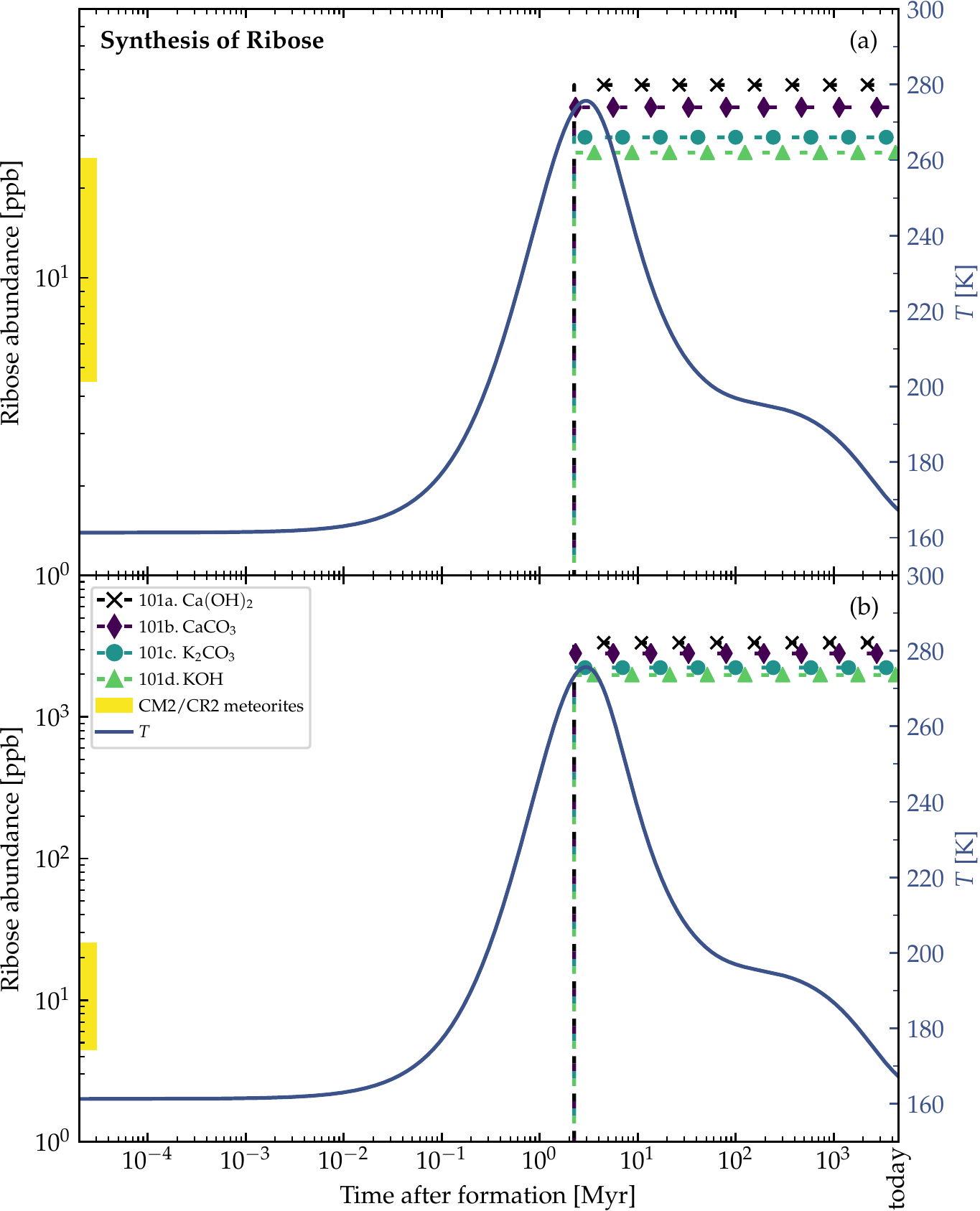}
    \caption{Theoretical ribose abundances in an outer shell at \SI{138}{\kilo\meter} distance from the center of the \SI{150}{\kilo\meter}-sized planetesimal model. The whole planetesimal model is shown in Figure~\ref{fig:planetesimal_150km}. Properties of planetesimal: ${\text{Radius} = \SI{150}{\kilo\meter}}$, densities ${\rho_{\mathrm{rock}} = \SI{3}{\gram\per\centi\meter\cubed}}$, ${\rho_{\mathrm{ice}} = \SI{0.917}{\gram\per\centi\meter\cubed}}$, porosity ${\phi = 0.2}$, and time of formation after ${\text{CAI} = \SI{3.5}{\mega\year}}$. The formose reaction pathway in Equation~\eqref{equ:formose} was used in the simulations. The experimentally found yields of ribose within 5Cs for each catalyst (see Table~\ref{tab:yields}) were multiplied with the theoretically calculated 5C abundance to obtain the ribose abundances (dashed lines with symbols). Ribose synthesis started at ${\sim}$$\SI{2.1}{\mega\year}$ after formation. All simulations were run at \SI{100}{bar}. In both panels (\textbf{a},\textbf{b}), the left vertical axis corresponds to the abundances (dashed lines with symbols) and the right vertical axis corresponds to the temperatures $T$ (solid lines) in the outer shell of the planetesimal model. The shaded part of the abundance axis represents the range of ribose abundances measured in CM2 (Murchison, upper limit) and CR2 (NWA 801, lower limit) meteorites \cite{Furukawa2019}, and has no correlation to the point in time (horizontal axis). (\textbf{a}) Time evolution of lower bound abundances simulated using the \textit{lower} (opposite to panel (\textbf{b})) bound of the initial concentration of glycolaldehyde of ${5\times10^{-6}\,\mathrm{mol}\cdot{}\mathrm{mol}_{\ce{H2O}}^{-1}}$ (see Table~\ref{tab:init_concs}). (\textbf{b}) Time evolution of upper bound abundances simulated using the \textit{upper} (opposite to panel (\textbf{a})) bound of the initial concentration of glycolaldehyde of ${4\times10^{-4}\,\mathrm{mol}\cdot{}\mathrm{mol}_{\ce{H2O}}^{-1}}$.\label{fig:shell_150km}}
\end{figure}

Figures~\ref{fig:shell_4km}~and~\ref{fig:shell_150km} show the ribose synthesis in the outer shells of the example planetesimal models in Figures~\ref{fig:planetesimal_4km}~and~\ref{fig:planetesimal_150km}. At the distance of \SI{2.76}{\kilo\meter} from the center of the \SI{4}{\kilo\meter}-sized planetesimal model (time of formation after ${\text{CAI} = \SI{1}{\mega\year}}$), the synthesis of ribose that started shortly before the peak temperature was reached in this region and allowed for liquid water to exist over $\lesssim${200,000} yr. When the water froze again, the formed ribose was preserved (see Figure~\ref{fig:shell_4km}). The similar phenomenon happened in the outer shell at the distance of \SI{138}{\kilo\meter} from the center of the \SI{150}{\kilo\meter}-sized planetesimal model (time of formation after ${\text{CAI} = \SI{3.5}{\mega\year}}$), where the water stayed liquid for $\lesssim\SI{2}{\mega\year}$ (see Figure~\ref{fig:shell_150km}).

It is worth noting that the planetesimal model by \citet{Lange2021}, which was used in our study, did not take into account the latent heat of water. Therefore, we were not able to accurately model the phase transitions of water and their effects on the temperature evolution. The phase transitions require a considerable amount of energy and cause the temperature evolution to stagnate until the water is completely melted or frozen. Since the periods estimated above were in between the melting and freezing times of water, our estimates were only approximations of the actual duration.

There should be a layer below the consistently frozen crust of planetesimal parent bodies, which could be the most promising part that contained the highest amounts of ribose. This region should be frozen rapidly due to the low and declining internal radioactive heating, preventing decomposition processes. On the other hand, the cores of the parent bodies were more likely to reach critically high temperatures over longer periods of time and probably led to significant sugar decomposition. Therefore, Figures~\ref{fig:R_lower}b~and~\ref{fig:R_upper}b could not depict the actual truth, as decomposition was not considered in our calculations.

The surface and outer shells of the parent bodies were more likely to be shattered and blown off by collisions with other asteroids. If impactors were able to penetrate through the surface layers of the parent bodies and reach the potentially ribose-rich intermediate part, the generated fragments would most likely contain the organic complexity that is characteristic for carbonaceous chondrites. This could explain why we were able to find ribose in carbonaceous chondrites on the Earth \cite{Furukawa2019}. The origin of these meteorites was likely biased to the outer shells of the parent bodies in fragmentation events, which could (partly) coincide with the most promising regions identified in Figures~\ref{fig:shell_4km}~and~\ref{fig:shell_150km}. 

\citet{Larralde1995} suspected that ribose was not stable enough to take part in the origin of life, questioning the RNA world hypothesis. However, ribose could be preserved in the icy fragments for a long time. This illustrates a possible explanation for how the concerns about the instability of ribose could be resolved in the scope of the chemical synthesis in planetesimals. By eventually falling as meteorites to the Earth into WLPs, the ribose-rich fragments might provide ribose for the origin of life, as described by \mbox{\citet{Pearce2017}}. 

Smaller planetesimals, such as the \SI{4}{\kilo\meter}-sized one in Figures~\ref{fig:planetesimal_4km}~and~\ref{fig:shell_4km}, had to be formed earlier than larger planetesimals to become aqueous, and allow for the formose reaction to take place in its interior. At \SI{1}{\mega\year} after CAI, there was enough \ce{^{26}Al} left to even heat the outer shells of the \SI{4}{\kilo\meter}-sized planetesimal above the melting point of water (see Figure~\ref{fig:shell_4km}). The large \SI{150}{\kilo\meter}-sized planetesimal was formed later at \SI{3.5}{\mega\year} after CAI (see Figures~\ref{fig:planetesimal_150km},~\ref{fig:R_lower},~\ref{fig:R_upper}~and~\ref{fig:shell_150km}). If it was formed earlier as the smaller planetesimal, this large body would have reached such high temperatures that strong thermal metamorphism or even siliceous volcanism would have occurred, resulting in hostile conditions for the synthesis of organic molecules. Therefore, when comparing these moderately heated planetesimals, the time intervals of the aqueous phase in the outer shells occurred earlier in the smaller bodies and over a shorter interval than in the larger ones ($\sim${200,000} yr in Figure~\ref{fig:shell_4km} vs. $\sim$$\SI{2}{\mega\year}$ in Figure~\ref{fig:shell_150km}). This leads to the conclusion that smaller planetesimals with moderate heating might preserve ribose better than larger planetesimals since the aqueous time interval allowing for decomposition of the formed ribose was shorter by around one order of magnitude.

In follow-up studies, when more detailed thermodynamic and kinetic data with a higher temperature resolution for the decomposition rates become available, it could be interesting to consider the decomposition of ribose in more detail and constrain the region with the likely highest ribose abundances with more accuracy compared to our approximate estimates. This could also help to identify the part of parent bodies where carbonaceous chondrites with high ribose content could have most likely originated.

In this study, we used the same model and the same initial concentrations of reactants as in our previous study for nucleobases \cite{Paschek2021}, in which we found abundances matching the measured values in meteorites. It seems that the formation of crucial RNA-building blocks such as nucleobases and ribose could be explained uniformly with our model and the selected reaction pathways. Note that carbonaceous chondrites also contain \ce{P}-rich minerals, e.g., schreibersite, which could provide the last missing piece for the synthesis of the RNA nucleotides, the phosphates (\ce{[PO4]^{3-}}, \ce{[HPO4]^{2-}}, and \ce{[H2PO4]-} depending on pH) \cite{Gull2015}. Moreover, the clay minerals at the bottom of WLPs could also provide the phosphates needed for the phosphorylation of nucleosides \cite{Ferris1996}. In addition, metal-doped-clays were shown to select ribose from a formose mixture \cite{Zhao2021} and catalyze the formation of ribonucleosides \cite{Chen2021}.

Thus, ribose and nucleobases (see our previous studies \cite{Pearce2016,Paschek2021}) delivered by carbonaceous chondrites could have been an essential ingredient for the build-up of the first RNA molecules in WLPs (including geothermal fields and hot springs), or around subsea hydrothermal vents, or all of them, setting the stage for the emergence of the RNA world and the origin of life on the Earth and elsewhere.

\newpage



\authorcontributions{Conceptualization, D.A.S.; methodology, K.P., K.K., B.K.D.P., K.L., O.T., R.E.P. and D.A.S.; software, K.P., B.K.D.P. and K.L.; validation, K.P., K.K., B.K.D.P., K.L., T.K.H. and D.A.S.; formal analysis, K.P., K.K., B.K.D.P. and K.L.; investigation, K.P., K.K., B.K.D.P., K.L. and D.A.S.; resources, T.K.H., O.T., R.E.P. and D.A.S.; data curation, K.P., K.K., B.K.D.P. and K.L.; writing---original draft preparation, K.P.; writing---review and editing, K.K., B.K.D.P., K.L., T.K.H., O.T., R.E.P. and D.A.S.; visualization, K.P.; supervision, T.K.H., O.T., R.E.P. and D.A.S.; project administration, T.K.H. and D.A.S.; funding acquisition, T.K.H. All authors have read and agreed to the published version of the manuscript.}

\funding{K.P.\ acknowledges financial support by the Deutsche Forschungsgemeinschaft (DFG, German Research Foundation) under Germany's Excellence Strategy EXC 2181/1-390900948 (the Heidelberg STRUCTURES Excellence Cluster).
K.P.\ is a fellow of the International Max Planck Research School for Astronomy and Cosmic Physics at the University of Heidelberg (IMPRS-HD).
T.K.H.\ acknowledges financial support by the European Research Council under the Horizon 2020 Framework Program via the ERC Advanced Grant Origins 83 24 28.
D.A.S.\ acknowledges financial support by the Deutsche Forschungsgemeinschaft through SPP 1833: ``Building a Habitable Earth'' (SE 1962/6-1).}

\institutionalreview{Not applicable.}

\informedconsent{Not applicable.}

\dataavailability{The source code, excluding the proprietary ChemApp library, and including the data of the planetesimal models, is openly available on Zenodo at (\cite{klaus_paschek_2021_5774880}, \url{https://doi.org/10.5281/zenodo.5774880}, accessed on 1 March 2022) and as a Git repository: \url{https://github.com/klauspaschek/prebiotic_synthesis_planetesimal}, accessed on 17 November 2021.}

\acknowledgments{The authors thank Cornelis P.~Dullemond for his extensive help in understanding the planetesimal model and the background attached to its theory. We would also like to thank Catharina Fairchild and Hao-En Huang for their stylistic review of the manuscript. We wish to thank an anonymous reviewer for corrections, comments, and suggestions.}

\conflictsofinterest{The authors declare no conflict of interest.} 



\abbreviations{Abbreviations}{
The following abbreviations are used in this manuscript:\\

\noindent 
\begin{tabular}{@{}ll}
RNA & Ribonucleic acid\\
DNA & Deoxyribonucleic acid\\
5C & Pentose\\
WLP & Warm little pond/Darwinian pond\\
ppb & Parts per billion\\
ppm & Parts per million\\
CAI & Calcium-aluminium-rich inclusions\\
yr & Year(s)\\
GC-MS & gas chromatography coupled with a mass spectrometer\\
CM2 & Mighei-type chondrites (CM), a group of meteorites, in this case of petrologic type 2\\
CR2 & Renazzo-type chondrites (CR), a group of meteorites, in this case of petrologic type 2
\end{tabular}}




\newpage
\begin{adjustwidth}{-\extralength}{0cm}
\reftitle{References}


\end{adjustwidth}

\end{document}